\documentclass[12pt]{article} \topmargin=0.5 cm  \textwidth 16.5
cm \textheight 23 cm \oddsidemargin 0pt \evensidemargin 0pt
\usepackage {graphicx}

\begin{document}

\title{Phase
shift approach in the case of a magnetic \\ flux tube}
\author{Pavlos Pasipoularides  \footnote{paul@central.ntua.gr} \\
       Department of Physics, National Technical University of
       Athens \\ Zografou Campus, 157 80 Athens, Greece}
\date{ }
       \maketitle

\begin{abstract}
In Refs. \cite{3,1:wg} two different approaches have been developed
for the numerical computation of the effective energy in the
presence of a magnetic flux tube, by using the phase shift method.
However, the opinion has been expressed that these two variations of
the phase shift method are not equivalent. In this paper we aim to
solve this ambiguity by comparing these two different approaches and
showing that they give identical numerical results, within the
numerical accuracy of the method.
\end{abstract}

\section{Introduction}

The numerical computation of the fermion-induced effective energy
\footnote{The effective energy can be written as
$E_{eff}=E_{clas}+E_{Q}$, where $E_{clas}$ is the classical energy
and $E_{Q}$ is the quantum part of the fermion induced effective
energy.}, in the presence of inhomogeneous magnetic fields of the
form of a flux tube, is a topic that has attract the attention of
several authors, see Refs. \cite{1,3,4,1:wg}.

In Refs. \cite{3,1:wg} two different approaches have been developed
for the numerical computation of the quantum energy in the presence
of a magnetic flux tube, by using the phase shift method. The
question that arises is whether these two different approaches are
equivalent or not. In this paper we show, mainly by performing
numerical computations, the equivalence of these two different
approaches. In this way we give a detailed answer to the objections
in Ref. \cite{1:wg}, according to which the integral \footnote{of
Eq. (26) in Ref. \cite{3} or in Eq. (12) in this work} we used for
the numerical computation of the 3+1 effective energy is not
properly renormalized, and as a result it does not give the correct
values for the effective energy.

For this reason, in section 2 we present in a detailed way the
renormalization procedure we followed in Ref. \cite{3}, and we
emphasize that exactly the same method has been used by other
authors, and especially in Ref. \cite{9}. In section 3 we present
our numerical results for the integrand of Eq. (12) for the quantum
energy, in the case of the Gaussian magnetic flux tube (of Eq. (16)
below). These results indicate a highly convergent behavior for the
integral of Eq. (12). In section 4, we compare our results (Fig.
\ref{2} in this work), for the quantum energy in the presence of the
Gaussian flux tube, with the correspondent results of Ref.
\cite{1:wg} (Fig. 4 in Ref. \cite{1:wg}), and we obtain that they
are identical, within the numerical error of the method. Finally, in
section 5 we compare the two methods more carefully by using a
different way (comparing with the vacuum polarization diagram) and
we find again that they should be equivalent.

\section{Renormalization of the quantum energy}

In this section we explain in detail the renormalization procedure
we used in Ref. \cite{3}.

The quantum part of the fermion-induced effective energy in the
presence of a magnetic field is given by the equation:
\begin{eqnarray}
E_{Q}=-\frac{1}{2iT}Tr\ln(\not\!\!D^{2}+m^{2}_{f})
\end{eqnarray}
where  $\not\!\!\!D=\gamma^{\mu}D_{\mu}$ $(\mu=0,1,2,3)$ and
$D_{\mu}=\partial_{\mu}-ieA_{\mu}$. The gamma matrices satisfy the
relationship $\{\gamma^{\mu},\gamma^{\nu}\}=2g^{\mu\nu}$, and $T$ is
the total length of time.

We have shown in Ref. \cite{3} that the $2+1$ quantum energy for a
magnetic field of the form of a flux tube is given by the equation
\begin{eqnarray}
E_{Q(2+1)}=\frac{1}{2\pi}\int_{0}^{+\infty}dk\frac{k}{\sqrt{k^{2}+m^{2}_{f}}}
(\Delta(k)-c)
\end{eqnarray}
where $c=\lim_{k \rightarrow +\infty}\Delta(k)$. Our numerical
study, in Ref. \cite{3}, shows that $c=-\pi \phi^{2}$ ($\phi=e
\Phi/2 \pi$). This means that $c$ is independent of the special form
of the magnetic field we examine and depends only on the total
magnetic flux of the field $\Phi$. We also note that $c$ arises from
an integration by parts (for details see Ref. \cite{3}) and it is
not a quantity we subtract and add back in order to make the above
integral convergent.

The function $\Delta(k)$ was defined in Ref. \cite{3} as the
symmetrical sum over l:
\begin{equation}
\Delta(k)=\lim_{L\rightarrow+\infty}\sum_{s,l=-L}^{L}\delta_{l,s}(k)
\end{equation}
where $\delta_{l,s}(k)$ is the phase shift which corresponds to
$l^{th}$ partial wave with momentum $k$ and spin $s$.

It is well known that there are no divergencies in 2+1 dimensional
QED. From this point of view we expect the above integral of Eq. (2)
to be convergent. In addition, our numerical results, for the
integrand of Eq. (2), confirm with very good accuracy this expected
convergent behavior, see Fig. 1 in Ref. \cite{3}, or Fig. \ref{1}
and Fig. \ref{3} below in this paper. The calculation of the phase
shifts have been performed by solving an ordinary differential
equation. For details see Refs. \cite{3,phase}.

The corresponding $3+1$ dimensional result is obtained by the $2+1$
quantum energy if we replace $m_{f}^{2}$ by $k_{3}^{2}+m_{f}^{2}$ in
(2) and integrate over the $k_{3}$ momentum \cite{9}. An overall
factor of $2$ (from the Dirac trace) must also be included.
\begin{eqnarray}
E_{Q(3+1)}=\frac{1}{\pi}\int_{0}^{+\infty}dk\int_{-\Lambda/2}^{\Lambda/2}\frac{dk_{3}}{2\pi/L_{z}}\frac{k}{\sqrt{k^{2}+k_{3}^2+m^{2}_{f}}}
(\Delta(k)-c)
\end{eqnarray}
As the integral over $k_{3}$ is divergent we have introduced a
regularization parameter $\Lambda/2$. Now, if we integrate over
$k_{3}$ we obtain the unrenormalized quantum energy per unit length
$L_{z}$ ($L_{z}$ is the length of the space box towards $z$
direction) equal to
\begin{eqnarray}
E_{Q(3+1)}=-\frac{1}{2\pi^{2}}\int_{0}^{+\infty} dk
k\ln\left(\frac{k^{2}+m^{2}_{f}}{m_{f}^{2}}\right)
(\Delta(k)-c)+\frac{1}{2
\pi^{2}}\ln(\frac{\Lambda^{2}}{m_{f}^{2}})\int_{0}^{+\infty} dk k
(\Delta(k)-c)
\end{eqnarray}

In the weak field limit $eB<<m_{f}^{2}$, or for large $m_{f}$, the
above expression for the effective energy must tend to the first
diagram (vacuum polarization diagram) of the perturbative expansion
of the effective energy:
\begin{eqnarray}
E_{Q(3+1)}^{(2)}=&-& \frac{1}{8\pi^{3}}\int_{0}^{1} dx x (1-x)
\int_{0}^{+\infty}dq q |\widetilde{B}(q)|^{2}
\ln(\frac{m_{f}^{2}+q^{2}x (1-x)}{m_{f}^{2}}) \nonumber \\
&+&\frac{1}{24\pi^{2}}\ln(\frac{\Lambda^{2}}{m_{f}^{2}})\int
d^{2}\vec{x} B^{2}(\vec{x})
\end{eqnarray}
where $\widetilde{B}(q)=\int
d^{2}\vec{x}\:e^{-i\vec{q}\:\cdot\vec{x}}B(\vec{x})$, and we have
assumed the rescaling $ B \rightarrow eB$.

Note that the vacuum polarization diagram for the 3+1 dimensional
case (see Eq. (6) above) can be obtained from 2+1 dimensional result
\begin{eqnarray}
E^{(2)}_{Q(2+1)}=\frac{1}{16\pi^{3}}\int
d^{2}\vec{q}\;|\widetilde{B}(\vec{q})|^{2}\int_{0}^{1}dx\frac{x(1-x)}{\sqrt{m_{f}^{2}+x(1-x)\vec{q}\:^{2}}}
\end{eqnarray}
if we set $m_{f}^{2}\rightarrow k_{3}^{2}+m_{f}^{2}$ and integrate
over the $k_{3}$ momentum (an overall factor of $2$ must also be
included).

By comparing Eqs. (5) and (6) in the weak field limit we obtain that
\begin{equation}
\int_{0}^{+\infty} dk k (\Delta(k)-c)=\frac{1}{12}\int d^{2}\vec{x}
B^{2}(\vec{x})
\end{equation}
Note that we have confirmed the above equation numerically by
performing numerical computation for several magnetic fields of the
form of a flux tube.

Now Eq. (5), for the unrenormalized quantum energy, can be written
as
\begin{eqnarray}
E_{Q(3+1)}&=&\frac{1}{2\pi^{2}}\int_{0}^{+\infty} dk
k\ln\left(\frac{k^{2}+m^{2}_{f}}{m_{f}^{2}}\right)
(\Delta(k)-c)\nonumber \\
&+&\frac{1}{24\pi^{2}}\ln(\frac{\Lambda^{2}}{m_{f}^{2}})\int
d^{2}\vec{x} B^{2}(\vec{x})
\end{eqnarray}
The logarithmically divergent term can be incorporated in the
classical energy, as is shown below
\begin{eqnarray}
E_{eff}&=&E_{class}+E_{Q} \nonumber\\
&=&\frac{1}{2}
\left(\frac{1}{e^{2}}+\frac{1}{12\pi^{2}}\ln(\frac{\Lambda^{2}}{m_{f}^{2}})\right)\int
d^{2}\vec{x} B^{2}(\vec{x}) \nonumber \\ &+&
\frac{1}{2\pi^{2}}\int_{0}^{+\infty} dk
k\ln\left(\frac{k^{2}+m^{2}_{f}}{m_{f}^{2}}\right) (\Delta(k)-c)
\end{eqnarray}
In this way the logarithmically divergent part corresponds to a
charge renormalization according to the equation
\begin{equation}
\frac{1}{e^{2}_{R}}=\frac{1}{e^{2}}+\frac{1}{12\pi^{2}}\ln(\frac{\Lambda^{2}}{m_{f}^{2}})
\end{equation}
and thus the renormalized quantum energy is
\begin{equation}
E_{Q(3+1)}^{(ren)}=-\frac{1}{2\pi^{2}}\int_{0}^{+\infty}dk
k\ln\left(\frac{k^{2}+m^{2}_{f}}{m_{f}^{2}}\right) (\Delta(k)-c)
\end{equation}
The above presented renormalization procedure corresponds to the
on-shell renormalization condition $\Pi(0)=0$ (for more details see
Ref. \cite{pesk}).

\textit{Now, as the divergent part has been removed, the integral of
Eq. (12), for the renormalized quantum energy, is expected to be
convergent.} In the next section, we will indeed see that our
numerical results indicate a highly convergent behavior for this
integral.

Note that exactly the same renormalization procedure has been used
by the authors of Ref. \cite{9}. Particularly, in Ref. \cite{9} the
analytical result for the 2+1 quantum energy for a special form of
an inhomogeneous magnetic field is taken for granted  from their
previous work of Ref \cite{Dun}. Then the replacement
$m_{f}^{2}\rightarrow k_{3}^{2}+m_{f}^{2}$ is performed, and the 3+1
dimensional result is obtained by integrating out the $k_{3}$
momentum. As we did in this paper, a cut-off $\Lambda/2$ is used in
Ref. \cite{9} for the regularization of the integral over $k_{3}$,
and the same logarithmically divergent term arises. This term can be
removed as it corresponds to a charge renormalization. In Ref.
\cite{9} we see that the remaining part, which corresponds to the
regularized quantum energy, is an analytical formula with no
divergencies.

\section{Convergence of the integral of Eq. (12)}

For the numerical computations we have chosen the Gaussian magnetic
field
\begin{equation}
B_{G}(r)=\frac{2\phi}{d^{2}}\: \exp(-\frac{r^{2}}{d^{2}})
\end{equation}
where $d$ is the spatial size of the magnetic flux tube,
$\phi=e\Phi/2\pi$, and $\Phi$ is the total magnetic flux of the
field.

\begin{figure}[h]
\begin{center}
\includegraphics[scale=1.5,angle=0]{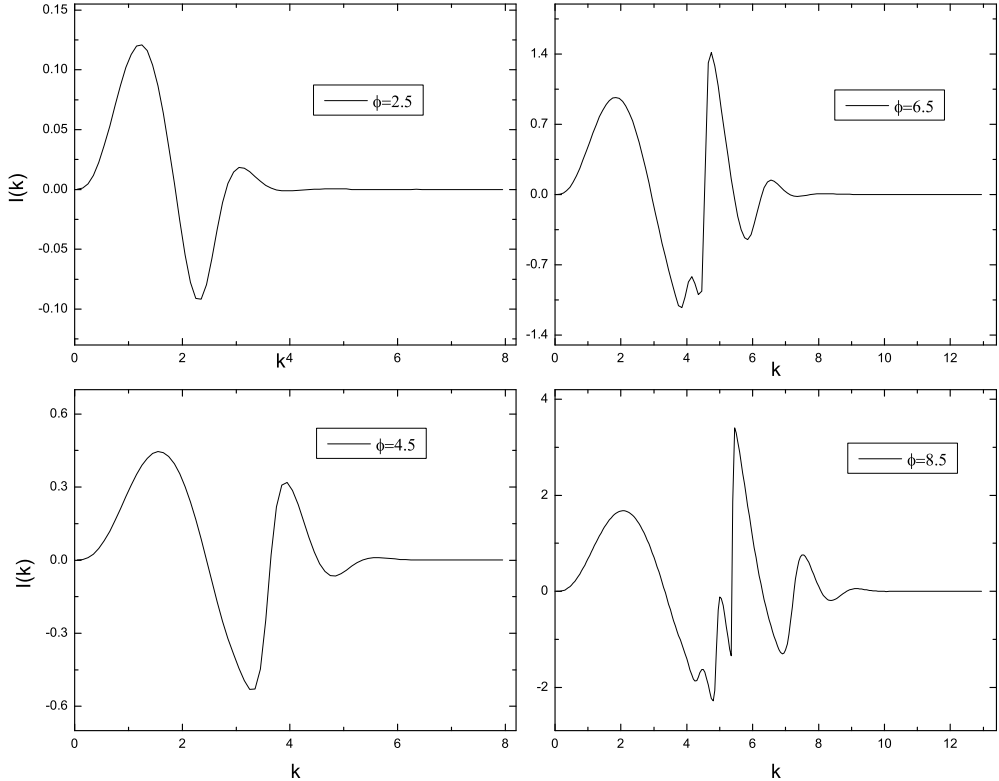}
\end{center}
\caption {The integrand $I(k)$ as a function of $k$ for
$\phi=2.5,4.5,6.5,8.5$ and $m_{f}=1$, in the case of the Gaussian
magnetic field of Eq. (16). } \label{1}
\end{figure}

In Fig. \ref{1} we have plotted the integrand of Eq. (12)
\begin{equation}
I(k)=\frac{1}{2\pi^{2}}
k\ln\left(\frac{k^{2}+m^{2}_{f}}{m_{f}^{2}}\right) (\Delta(k)-c)
\end{equation}
as a function of $k$ for $\phi=2.5,4.5,6.5,8.5$, $d=1$ and
$m_{f}=1$.

We see that our numerical results in Fig. \ref{1} indicate a highly
convergent behavior for the integral of Eq. (12), as the integrand
$I(k)$ becomes zero for sufficiently large values of $k$.

\section{Comparing with the results of Ref. \cite{1:wg}}

\begin{figure}[h]
\begin{center}
\includegraphics[scale=1.7,angle=0]{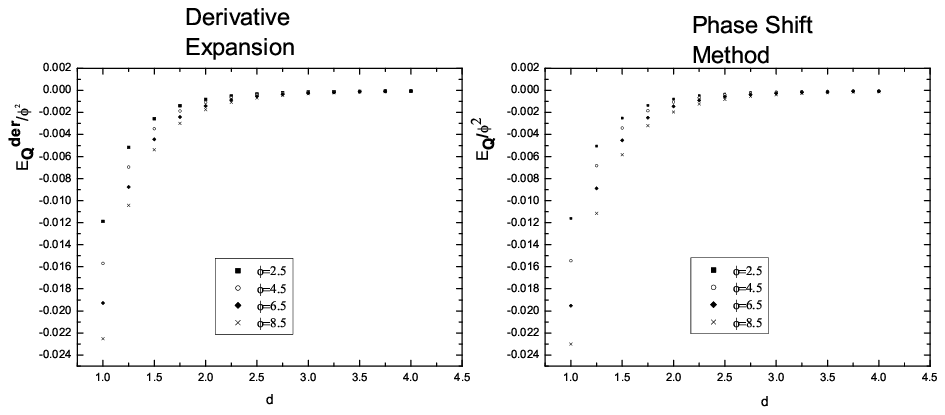}
\end{center}
\caption {Derivative Expansion versus Phase Shift Method for the
Gaussian magnetic flux tube, for $\phi=2.5,4.5,6.5,8.5$ and
$m_{f}=1$.} \label{2}
\end{figure}

As the convergence of the integral of Eq. (12) is not sufficient to
show that our numerical results for the quantum energy are the same
with those of the phase shift approach of Ref. \cite{1:wg}, we
should perform a comparison of the numerical results of these two
different approaches.

For this we have plotted $E^{(ren)}_{Q(3+1)}/\phi^{2}$ as a function
of $d$ for $\phi=2.5,4.5,6.5,8.5$ and $m_{f}=1$, in the right-hand
panel of Fig. \ref{2}. We see that our results in this figure  are
identical (within the numerical accuracy of the method) with the
corresponding results of Ref. \cite{1:wg} (\textit{see the
right-hand panel of Fig. (4) in Ref. \cite{1:wg}}).

Also, it is worth to compare the results of the phase shift method
with the derivative expansion, which is the standard approximative
tool in the case of smooth inhomogeneous magnetic fields (see for
example Ref. \cite{der5}). In Fig. \ref{2} we see that our results
are in a very good agreement with the corresponding results of the
derivative expansion in the homogeneous limit $1/\sqrt{B_{m}}<<d$,
as is expected, where $B_{m}=2 \phi/d^{2}$ .

\section{Phase shift approach and vacuum polarization diagram}

\begin{figure}[h]
\begin{center}
\includegraphics[scale=1.2,angle=0]{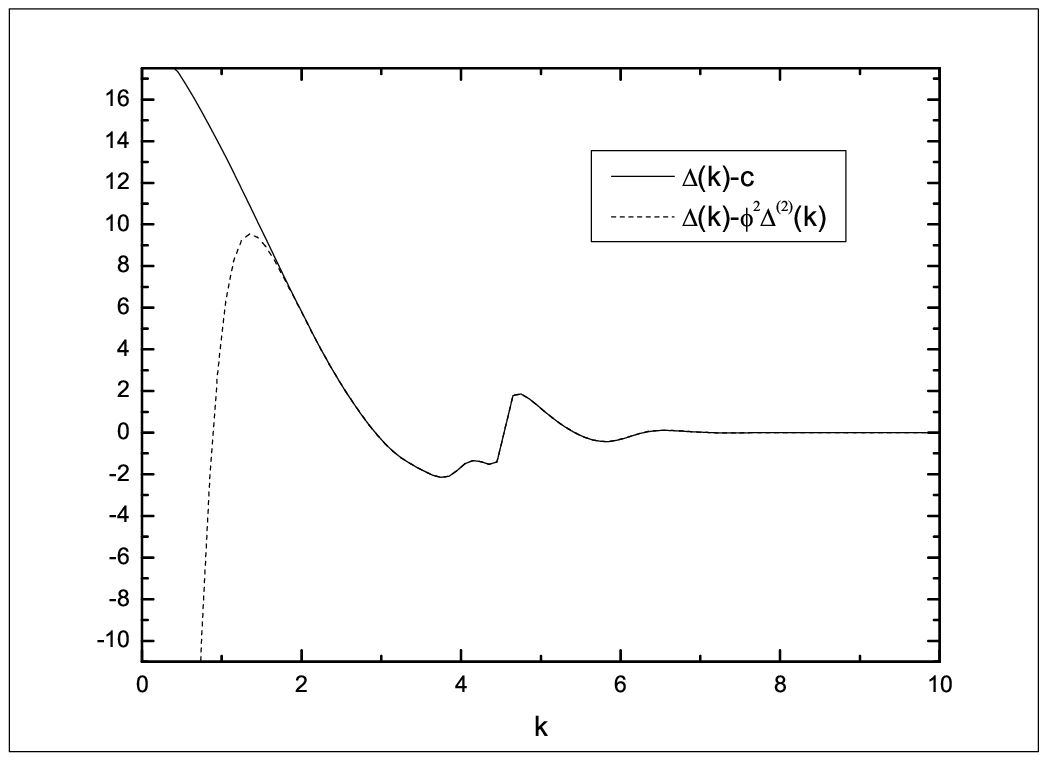}
\end{center}
\caption {$\Delta(k)-c$ and
$\overline{\Delta}(k)=\Delta(k)-\phi^{2}\Delta^{(2)}(k)$ as a
function of k for $\phi=6.5$, $m_{f}=1$ and d=1, where $c=-\pi
\phi^{2}$. } \label{3}
\end{figure}

\begin{figure}[h]
\begin{center}
\includegraphics[scale=1.2,angle=0]{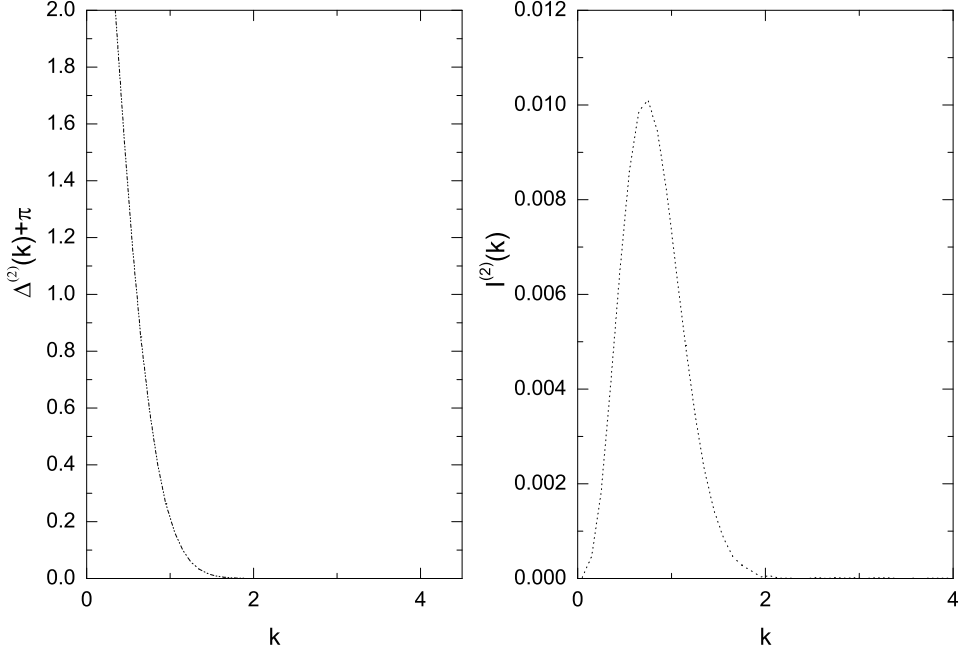}
\end{center}
\caption {$\Delta^{(2)}(k)+\pi$ and $I_{2}(k)$ as a function of k
for d=1 and $m_{f}=1$. We see that $\Delta^{(2)}(k)\rightarrow-\pi$
as $k\rightarrow +\infty$.} \label{4}
\end{figure}

\begin{figure}[h]
\begin{center}
\includegraphics[scale=1.2,angle=0]{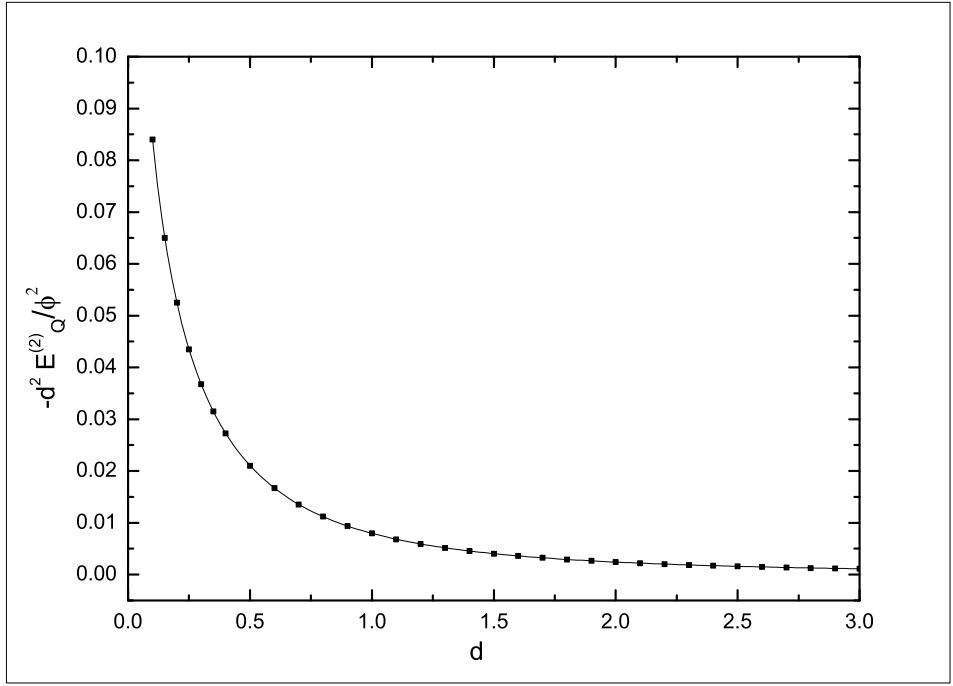}
\end{center}
\caption {We have plotted the first and the second member of Eq.
(21), multiplied by $-d^{2}/\phi^{2}$, as a function of $d$ for
$m_{f}=1$. The continuous line corresponds to the vacuum
polarization diagram, and the discrete points to the phase shift
integral. According to this figure, Eq. (21) is valid.} \label{5}
\end{figure}

In Ref. \cite{1:wg} the renormalized quantum energy \footnote{In
what will follow we have dropped the index (ren) from the
renormalized quantum energy $E^{(ren)}_{Q(3+1)}$.} $E'_{Q(3+1)}$ is
computed as a sum of two terms
\begin{equation}
E'_{Q(3+1)}=E_{ph(3+1)}+E^{(2)}_{Q(3+1)}
\end{equation}
where the first term corresponds to the contribution of the phase
shifts
\begin{equation}
E_{ph(3+1)}=-\frac{1}{2\pi^{2}}\int_{0}^{+\infty}dk
k\ln\left(\frac{k^{2}+m^{2}_{f}}{m_{f}^{2}}\right)
\overline{\Delta}(k)
\end{equation}
and the second term $E^{(2)}_{Q(3+1)}$ is the renormalized quantum
energy which corresponds to the vacuum polarization diagram (see Eq.
(8)).

The function $\overline{\Delta}(k)$ is defined as
\begin{equation}
\overline{\Delta}(k)=\lim_{L\rightarrow+\infty}\sum_{s,l=-L}^{L}\overline{\delta}_{l,s}(k)
\end{equation}
where
\begin{equation}
\overline{\delta}_{l,s}(k)={\delta}_{l,s}(k)-\phi
{\delta}_{l,s}^{(1)}(k)-\phi^{2} {\delta}_{l,s}^{(2)}(k)
\end{equation}
and ${\delta}_{l,s}^{(1)}(k)$, ${\delta}_{l,s}^{(2)}(k)$ are the
coefficients of $\phi$ and $\phi^{2}$ which arise if we expand the
phase shift ${\delta}_{l,s}(k)$ in powers of $\phi$.

We emphasize that ${\delta}_{l,s}^{(1)}(k)$ and
${\delta}_{l,s}^{(2)}(k)$ are not the first and second Born
approximations, as the Born expansion is an expansion in powers of
the potential $v_{l,s}(r)$ \footnote{For the exact form of the
potential in the presence of the flux tube see Eq. (17) in Ref.
\cite{3}} of the corresponding scattering problem, and this
potential includes both terms of $\phi$ and $\phi^{2}$.

From Eqs. (17) and (18) we obtain
\begin{equation}
\overline{\Delta}(k)=\Delta(k)-\phi^{2}\Delta^{(2)}(k)
\end{equation}
as we can prove, from the integral formula for the first Born
approximation, that $\Delta^{(1)}(k)=0$.

Note that the functions $\Delta^{(1)}(k)$ and $\Delta^{(2)}(k)$ are
defined as:
\begin{eqnarray}
\Delta^{(1)}(k)=\lim_{L\rightarrow+\infty}\sum_{s,l=-L}^{L}\delta^{(1)}_{l,s}(k)\;,
\qquad
\Delta^{(2)}(k)=\lim_{L\rightarrow+\infty}\sum_{s,l=-L}^{L}\delta^{(2)}_{l,s}(k)
\nonumber
\end{eqnarray}

In Fig. \ref{3} we have plotted $\Delta(k)-c=\Delta(k)+\pi \phi^{2}$
and $\overline{\Delta}(k)=\Delta(k)-\phi^{2}\Delta^{(2)}(k)$ as a
function of k for $\phi=6.5$, $m_{f}=1$ and d=1. We see that very
shortly ($k>2$) these two functions become identical. The main
reason is that the function $\Delta^{(2)}(k)$ tents rapidly to
$-\pi$, as we see in the left-hand panel of Fig. \ref{4}.

The immediate consequence of the above discussion is that the
integrals of Eqs. (12) and (16) have exactly the same convergent
properties. Thus, Eq. (16), which was used for the computation of
the quantum energy in Ref. \cite{1:wg}, has no additional advantage
against Eq. (12) which we used in our work of Ref. \cite{3}. On the
other hand, if we use Eq. (12) for the quantum energy, it is not
necessary to compute additional quantities like the phase shifts
${\delta}_{l,s}^{(2)}(k)$ and the vacuum polarization diagram.

In addition, in the previous section we see that the results of
these two variations of the phase shift method are in a very good
agreement, however in this section we will go over this topic more
carefully.

If we demand the results of two methods to be equal
($E_{Q(3+1)}=E'_{Q(3+1)}$ see Eqs (12) and (15)), from Eqs. (6),
(12), (15), (16) we obtain
\begin{eqnarray}
&-&\frac{1}{2\pi^{2}}\int_{0}^{+\infty}dk
k\ln\left(\frac{k^{2}+m^{2}_{f}}{m_{f}^{2}}\right)
\left(\Delta(k)-c\right)=-\frac{1}{2\pi^{2}}\int_{0}^{+\infty}dk
k\ln\left(\frac{k^{2}+m^{2}_{f}}{m_{f}^{2}}\right)
\overline{\Delta}(k)\nonumber \\&-&\frac{1}{8\pi^{3}}\int_{0}^{1} dx
x (1-x) \int_{0}^{+\infty}dq q |\widetilde{B}(q)|^{2}
\ln(\frac{m_{f}^{2}+q^{2}x (1-x)}{m_{f}^{2}})
\end{eqnarray}
or, if we use Eq. (19), we obtain the equivalent equation
\begin{eqnarray}
&-&\phi^{2}\frac{1}{2\pi^{2}}\int_{0}^{+\infty}dk
k\ln\left(\frac{k^{2}+m^{2}_{f}}{m_{f}^{2}}\right)
\left(\Delta^{(2)}(k)+\pi\right)\nonumber
\\&=&-\frac{1}{8\pi^{3}}\int_{0}^{1} dx x (1-x) \int_{0}^{+\infty}dq
q |\widetilde{B}(q)|^{2} \ln(\frac{m_{f}^{2}+q^{2}x
(1-x)}{m_{f}^{2}})
\end{eqnarray}
In Fig. \ref{5} we have plotted the first and the second member of
Eq. (21), multiplied by $-d^{2}/\phi^{2}$, as a function of $d$ for
$m_{f}=1$. The continuous line corresponds to the vacuum
polarization diagram, and the discrete points to the phase shift
integral of the first member of Eq. (21). Thus, from Fig. \ref{5},
it is seen that the first and second member of Eq. (21) are equal
(the deviations between the numerical values of the first and second
member of Eq. (21) are of the order of 0.1 per cent). As a result
the two alternative approaches, for the quantum energy in the
presence of a flux tube, are equivalent.

In addition, if take into account the above results, we can compute
the difference $|E_{Q(3+1)}-E'_{Q(3+1)}|$, where $E_{Q(3+1)}$ and
$E'_{Q(3+1)}$ are the renormalized quantum energies which correspond
to the two variations of the phase shift method, see also Eq. (12)
and (16). We see that $|E_{Q(3+1)}-E'_{Q(3+1)}|/E_{Q(3+1)}\approx
0.001$, which is of the order of the accuracy of the numerical
computations, and as a result the numerical values for $E_{Q(3+1)}$
and $E'_{Q(3+1)}$ are identical.

Note that, according to the above discussion, the first member of
Eq. (21) can be viewed as the phase shift representation of the
vacuum polarization diagram, or we can write
\begin{equation}
E_{Q(3+1)}^{(2)}=-\phi^{2}\frac{1}{2\pi^{2}}\int_{0}^{+\infty}dk
k\ln\left(\frac{k^{2}+m^{2}_{f}}{m_{f}^{2}}\right)
\left(\Delta^{(2)}(k)+\pi\right)
\end{equation}

In order to study the convergent properties of the above integral,
we have plotted, in the right-hand panel of Fig. \ref{4}, the
integrand of Eq. (22)
\begin{equation}
I_{2}(k)=\frac{1}{2\pi^{2}}
k\ln\left(\frac{k^{2}+m^{2}_{f}}{m_{f}^{2}}\right)
\left(\Delta^{(2)}(k)+\pi\right)
\end{equation}
for $d=1$ and $m_{f}=1$. We see that that Fig. \ref{4} indicates a
highly convergent behavior for the integral of Eq. (22).

\section{Conclusions}

In this paper we compared the results of the two different
variations of the phase shift method, of Refs. \cite{3,1:wg}, for
the numerical computation of the quantum energy in the presence of a
magnetic flux tube. In section 3 and 4 we show, mainly by performing
numerical computations, that these two alternative approaches give
identical values for the quantum energy
($|E_{Q(3+1)}-E'_{Q(3+1)}|/E_{Q(3+1)}\approx 0.001$, see section 4),
and as a result they are equivalent.

In addition, in section 4 we show that in order to make the integral
over k, of Eq. (12), convergent, it is not necessary to subtract the
$\phi$ and $\phi^{2}$ parts of the phase shift $\delta_{l,s}(k)$
(see Eq. 18) and to add them back in their Feynman diagrammatic
form, as is done in Ref. \cite{1:wg}. This immediate convergence of
the integral of Eq. (12) is a consequence of the translation
invariance along the $x_{3}$ axis. However, in the case of problems,
e.g. with spherical symmetry, where the translation invariance is
violated, the subtraction of the asymptotic part of the phase shift
sum $\sum_{s,l=0}^{+\infty}\delta_{l,s}(k)$ (or the subtraction of
the first and second Born approximation), in order to make the
integral over $k$ convergent, is unavoidable (see Refs.
\cite{phase,den}).

It is worth to mention, that for the computation of the quantum
energy in the presence of a magnetic flux tube, two other
alternative methods have been developed: $a)$ the Jost function
method in Ref. \cite{1} and $b)$ the method of the worldline
numerics in Ref. \cite{4}. Note, that in Ref. \cite{3} we have
compared our results to the Jost function method, in the case of a
discontinuous magnetic field which is constant inside a cylinder of
radius $d$, and zero outside it. From Fig. 4 in Ref. \cite{3} and
the Fig. 3 in Ref. \cite{1} we see a very good agreement between the
phase shift and Jost function method.

Finally, it would be interesting if the two alternative approaches (
the Jost function and the worldline method) could give a figure like
that of the right-hand panel of Fig. \ref{2} of this work, for a
smooth realistic magnetic field like the Gaussian magnetic flux
tube.

\section{Acknowledgements}
I am grateful to Professors G. Tiktopoulos and K. Farakos for
important discussions. The work of P.P was supported by the
"Pythagoras" project of the Greek Ministry of Education.

\end{document}